\documentclass[prx,aps,twocolumn,showpacs,amsmath,amssymb,superscriptaddress,floatfix]{revtex4-2}

\usepackage{graphicx}
\usepackage[export]{adjustbox}
\usepackage{dcolumn}
\usepackage{upgreek}
\usepackage{bm,dsfont}
\usepackage{amssymb,mathtools}
\usepackage{booktabs}
\usepackage{color}
\usepackage{microtype}
\usepackage[T1]{fontenc}
\usepackage{lmodern} % for MiKTeX
\usepackage[shortlabels]{enumitem}
\usepackage{changes}
\usepackage{physics}
\usepackage{derivative}

\allowdisplaybreaks
\usepackage{placeins}
\usepackage[colorlinks,plainpages=false,pdfpagemode=UseNone,pdfstartview=FitBH]{hyperref}
\hypersetup{
colorlinks = true,
linkcolor = [rgb]{0.87,0.18,0.22},
citecolor = [rgb]{0.0,0.60,0.32},
urlcolor = [rgb]{0.20, 0.30, 0.54}}

\usepackage{sidecap}

\renewcommand{\tablename}{Tab.}
\makeatletter\renewcommand{\fnum@table}[1]{\tablename~\thetable.}\makeatother

\newcommand{\sro}{Sr$_2$RuO$_4$ }
\newcommand{\ophattext}[2]{\hat{#1}_{\text{#2} } }
\newcommand{\ophat}[2]{\hat{#1}_{#2} }

\newcommand{\estG}{g}
\definecolor{eggplant}{RGB}{180,33,147}

\include{tikzFig}

\definecolor{citecolor}{rgb}{0.0,0.60,0.32}
\newcommand{\CCQ}{Center for Computational Quantum Physics, Flatiron Institute, 162 5th Avenue, New York, NY 10010, USA}

\begin{document}

\title{Finite Temperature Minimal Entangled Typical Thermal States Impurity Solver}

\author{Xiaodong Cao}
\email{xcao@flatironinstitute.org}
\affiliation{\CCQ}

\author{E.\ Miles Stoudenmire}
\affiliation{\CCQ}

\author{Olivier Parcollet}
\affiliation{\CCQ}
\affiliation{Universit\'e Paris-Saclay, CNRS, CEA, Institut de Physique Th\'eorique, 91191, Gif-sur-Yvette, France}

%\date{\today}
\pacs{}
\begin{abstract}
We present a minimally entangled typical thermal state (METTS) quantum impurity
solver for general multi-orbital systems at finite temperatures. 
We introduce an improved estimator for the single-particle Green's function
that strongly reduces the large fluctuations at long imaginary time and low temperature,
which were a severe limitation of the original algorithm.
In combination with the fork tensor product states ansatz, we obtain 
a dynamical mean field theory (DMFT) quantum impurity solver, which 
we benchmark for single and three-band models down to low temperatures, 
including the effect of spin-orbit coupling in  
a realistic DMFT computation for the Hund's metal \sro down to low temperatures.
%As our method is insensitive to the fermionic sign problem, 
%it allows us to study the effect of spin-orbit coupling in  
%a realistic DMFT computation for the Hund's metal \sro down to low temperatures, a regime of parameter 
%inaccessible to quantum Monte Carlo methods.
\end{abstract}
\maketitle

\section{Introduction}\label{sec:intro}
Strongly correlated materials are a central topic in condensed matter physics.
Understanding or predicting their properties is a challenge
that requires the development of non-perturbative quantum
many-body methods. An example of such a method is the dynamical mean-field
theory (DMFT), which has significantly advanced
our understanding of these systems~\cite{Metzner1989,
Georges1992, Georges1996, Imada1998, 
Kotliar2004, Kotliar2006, Georges2013}. DMFT
maps the original lattice problem into an effective impurity problem with
a self-consistent bath. Solving such impurity problems 
remains a central computational bottleneck in the ab-initio study of strongly correlated material with DMFT.

The most common method for solving DMFT quantum impurity problems is quantum Monte Carlo
(QMC)~\cite{Gull2011, Georges1992, Ulmke1995, Rubtsov2005, Werner2006,
Werner2006b, Gull_2008}. However, its application is hindered by the intrinsic fermionic sign
problem at low temperatures in many interesting cases, e.g. systems with low symmetries or spin-orbit coupling. 
Exact diagonalization~\cite{Caffarel1994, Sangiovanni2006, Capone2007, Koch2008,
Zgid2012, Lin2013, Lu2014, AMARICCI2022108261} and numerical renormalization
group~\cite{Wilson1975, Bulla2008, Bulla1999, Bulla2011, Bulla2005,
Pruschke2000, kugler2020strongly, fabian2022}
are free from the sign problem but are
constrained to a small number of orbital degrees with high symmetries.
Recently, matrix product states (MPS)~\cite{White1992, White1993,
Schollwock2005, Schollwock2011} and its tree tensor network extensions have
emerged as successful many-body wave function ansatzes for general impurity
problems and have been successfully employed as impurity solvers on both real
and imaginary axis~\cite{Hallberg2006, Garcia2004, Raas2004, Holzner2010,
Ganahl2014, Ganahl2015, Wolf2014, Wolf2014b, Wolf2015,
Bauernfeind2017,linden2020,Cao2021}. However, these tensor network 
based impurity solvers are originally restricted to zero temperature
calculations.

Within the tensor network framework,
finite temperature computations can be conducted in two ways:
{\it i)} 
the minimal entangled typical thermal states
(METTS)~\cite{White2009, Miles2010, Barthel2017}, a Monte Carlo
sampling of states with low entanglement, and  {\it ii)} the purification
method~\cite{Feiguin2005, Zwolak2004, Verstraete2004, Hauschild2018}, which
accounts for both thermal and quantum fluctuations by doubling the Hilbert
space. Recently, a hybrid method was proposed that combines both
approaches~\cite{Jing2020,chung2019}.
It has been extended to the computation of the single-particle Green's function, in a single-orbital Anderson
impurity model~\cite{Bauernfeind2022}.
However in the algorithm presented in Ref.~\onlinecite{Bauernfeind2022}, the Green's function 
exhibits large variance in the METTS sampling at long imaginary time (i.e. for $\tau \sim \beta/2$),
which severely limits the capabilities of the method as an impurity solver at low frequencies.

In this work, we show how to reduce this variance significantly
using an improved estimator for Green's function.
Combined with the fork tensor product state
 many-body wave function ansatz,
we obtain an impurity solver for
general impurity problems at finite temperatures, which we show to be efficient for
three band systems, down to very low temperatures, even in the presence of spin-orbit coupling (SOC).

The paper is organized as follows. 
In Sec.~\ref{sec:method}, we introduce our improved estimator for the computation of the 
imaginary time Green's function in the hybrid METTS/purification method.
In Sec.~\ref{sec:result:improved_estimator}, we use single-band benchmarks to demonstrate the efficiency of the improved estimator with excellent agreement on the self-energy.
In  Sec.~\ref{sec:result:sro}, we apply our method to a realistic DMFT study of \sro and show the effect of SOC in parameter regimes that are inaccessible to QMC algorithms.
Finally, we conclude in Sec.~\ref{sec:conclusion}.
%Finally, other detailed benchmarks for three band systems %in regimes where the quantum Monte Carlo (CTHYB) is efficient 
%are presented in Appendix~\ref{app:threeband}.

%%%%%%%%%%%%%%%%%%%% METHOD %%%%%%%%%%%%%%%%%%%%%%%%%555

\section{METTS improved estimator for imaginary time Green's functions}\label{sec:method}
We consider a multiorbital Anderson impurity model coupled to a finite bath
defined by
\begin{align}\label{eq:anderson}
\ophattext{H}{} &= \ophattext{H}{loc} + \ophattext{H}{bath}, \\ \nonumber
\ophattext{H}{loc} &= \sum_{\alpha} \varepsilon_{\alpha_1\alpha_2}\ophat{d}{\alpha_1}^\dagger \ophat{d}{\alpha_2} + \sum_{\alpha} U_{\alpha_1\alpha_2\alpha_3\alpha_4} \ophat{d}{\alpha_1}^\dagger\ophat{d}{\alpha_2}^\dagger \ophat{d}{\alpha_3}\ophat{d}{\alpha_4}, \\ \nonumber
\ophattext{H}{bath} &= \sum_{i=1}^{N_b}
\left(
   \sum_{\kappa}\epsilon^i_{\kappa}\ophat{c}{i,\kappa}^\dagger\ophat{c}{i,\kappa} + \sum_{\kappa\alpha} 
%   \sum_{\kappa_1\kappa_2}\epsilon^i_{\kappa_1\kappa_2}\ophat{c}{i,\kappa_1}^\dagger\ophat{c}{i,\kappa_2} + \sum_{\kappa\alpha} 
   %\bigl( 
   V^i_{\alpha\kappa}\ophat{d}{\alpha}^\dagger \ophat{c}{i,\kappa} + {h.c.}
   %\bigr)
\right)
\end{align}
where $\ophattext{H}{bath}$ is the bath, 
$\ophattext{H}{loc}$ is the local impurity with its interactions,
including the Coulomb interaction $U_{\alpha_1\alpha_2\alpha_3\alpha_4}$ and various intra-site effects 
$\varepsilon_{\alpha_1\alpha_2}$.
Here, $\alpha$ (resp. $\kappa$) denotes the spin and orbital degrees of freedom of the impurity (resp. bath);
$i$ is the bath site index, and $N_b$ is the number of bath sites per spin-orbital.
In this Hamiltonian representation, the hybridization function reads
\begin{equation}\label{eq:finitebathDelta}
\Delta_{\alpha_1\alpha_2}(i\omega_n) = \sum_{i=1}^{N_b} \sum_{\kappa} \frac{V^{i}_{\alpha_1\kappa} V^{i*}_{\alpha_2\kappa}}{i\omega_n - \epsilon^{i}_{\kappa}},
%\Delta_{\alpha_1\alpha_2}(i\omega_n) = \sum_{i=1}^{N_b} \sum_{\kappa_1\kappa_2} \frac{V^{i}_{\alpha_1\kappa_1} V^{i*}_{\alpha_2\kappa_2}}{i\omega_n - \delta_{\kappa_1\kappa_2}\epsilon^{i}_{\kappa_1\kappa_2}},
\end{equation}
%using a diagonal on-site potential $\epsilon^{i}$.
In the context of DMFT computation, where the hybridization function $\Delta$ is the input of the impurity solver, 
$\Delta$ is fitted to the finite bath form of Eq.~\eqref{eq:finitebathDelta} using a standard procedure presented
in Appendix \ref{app:DeltaFit}.

\subsection{METTS and Purification}\label{sec:metts}
The central quantity in solving DMFT self-consistency equations is the single-particle Green's function, defined as
\begin{align}\label{eq:singl_particle}
G_{\alpha_1\alpha_2}(\tau) & = -\langle \mathcal{T}\ophat{d}{\alpha_1}(\tau) \ophat{d}{\alpha_2}^\dagger \rangle \\ \nonumber
&= -\frac{1}{\mathcal{Z}}\sum_{i}\bra{i} e^{-\beta\hat{H}} \mathcal{T}\ophat{d}{\alpha_1}(\tau) \ophat{d}{\alpha_2}^\dagger \ket{i},
%&= -\frac{1}{\mathcal{Z}}\sum_{i}\bra{i} e^{-\beta\hat{H}/2} \mathcal{T}\ophat{d}{\alpha_1}(\tau) \ophat{d}{\alpha_2}^\dagger e^{-\beta\hat{H}/2} \ket{i} \\ \nonumber
%&=-\frac{1}{\mathcal{Z}} \sum_{i} p(i)\bra{\phi_i} \mathcal{T}\ophat{d}{\alpha_1}(\tau) \ophat{d}{\alpha_2}^\dagger \ket{\phi_i},
\end{align}
where $\beta=1/T$ denotes the inverse temperature, $\mathcal{T}$ is the time
ordering operator and $\mathcal{Z} = \sum_{i}\bra{i} e^{-\beta\hat{H}}\ket{i}$
is the partition function. The states $\{\ket{i}\}$ represent a complete and
orthonormal basis. 
Using the cyclic property of the trace, the Green's function can be rewritten for $\tau >0$ as
\begin{align}
G_{\alpha_1\alpha_2}(\tau) & = -\frac{1}{\mathcal{Z}}\sum_{i}\bra{i} e^{-\beta\hat{H}/2} %\mathcal{T}
\ophat{d}{\alpha_1}(\tau) \ophat{d}{\alpha_2}^\dagger e^{-\beta\hat{H}/2} \ket{i}   \nonumber \\
& =-\frac{1}{\mathcal{Z}} \sum_{i} p(i)\bra{\phi_i} %\mathcal{T}
\ophat{d}{\alpha_1}(\tau) \ophat{d}{\alpha_2}^\dagger \ket{\phi_i},
\end{align}
defining the normalized states
$\ket{\phi_i}\equiv e^{-\beta\hat{H}/2}\ket{i}/\sqrt{p(i)}$ 
and the probability 
\mbox{$p(i)\equiv\bra{i}e^{-\beta\hat{H}}\ket{i}$}.

When $\{\ket{i}\}$ are chosen to be unentangled states, $\ket{\phi_i}$
are minimally entangled typical thermal states or ``METTS''
and $p(i)$ is the unnormalized probability weight of each METTS~\cite{White2009,Miles2010}.
In general, a METTS  $\ket{\phi_i}$ is a many-body state whose entanglement
grows smoothly from zero at smaller $\beta$.
For large $\beta$, each METTS approaches the ground state. METTS states can be importance sampled using an algorithm
where each METTS is ``collapsed'' into a product state that generates the next
METTS~\cite{White2009,Miles2010}. Thus the METTS approach is a quantum Monte
Carlo algorithm involving entangled rather than classical configurations.
%Then, $\ket{i}$ undergoes imaginary time evolution to $\beta/2$, producing the corresponding METTS $\ket{\phi_i}$. After measuring the Green's function defined in Eq.~\eqref{eq:singl_particle}, a new basis state $\ket{j}$ is selected with a transition probability of $p(i\rightarrow j) = |\langle j\ket{\phi_i}|^2$. One can demonstrate the detailed balance $p(i)p(i\rightarrow j) = p(j)p(j\rightarrow i)$, hence constructing a Markov chain importance sampling procedure that is free from sign problem~\cite{Miles2010}. 

To represent each METTS wavefunction, we adopt a tensor network with a 
``fork'' structure, where each interacting impurity site has its own
separate chain of bath degrees of freedom attached to it.  Such a fork tensor
product state ansatz has been demonstrated to efficiently capture the
entanglement structure of states of multi-orbital models~\cite{Bauernfeind2017,
Bauernfeind2018,Cao2021}.  To perform the imaginary time evolution necessary to
construct each METTS, we use the time-dependent variational principle (TDVP)
combined with a global basis
expansion~\cite{Haegeman2011,Haegeman2016,Bauernfeind2020,Cao2021,Mingru2020},
which has been demonstrated capable of providing highly accurate results for
impurity problems. Tensor operations are implemented using the ITensor
library~\cite{itensor,itensor-r0.3}. Further implementation details can be found in
Appendix~\ref{app:implementation_details}.

A possible drawback of the METTS approach is that it can require a large number
of samples to achieve desirable precision.  One method that has been proposed to
reduce the number of samples is to ``purify'' a subset of the sites of the
initial state generating each METTS~\cite{Jing2020,chung2019,Bauernfeind2022}. To purify a
site, one introduces a new corresponding ``ancilla'' site and prepares the purified site to be maximally entangled with the ancilla site before
time-evolving the state, i.e.,
\begin{align*}
&\ket{i} = \underset{x\in I_p}{\otimes} \left[\frac{1}{\sqrt{2}}\left( \ket{0}_{P_x}\ket{1}_{A_x} + \ket{1}_{P_x}\ket{0}_{A_x} \right)\right] \underset{y\not\in I_p}{\otimes}\ket{s_y},
\end{align*}
where $I_p$ represents the set of purified site indices. $P_x$ and $A_x$ subscripts indicate the physical and auxiliary degrees of freedom of site $x$, respectively. $\ket{s_y}\in\mathcal{H}_y\equiv \{ \ket{0}_y, \ket{1}_y \}$ is a state in the local Hilbert space of site $y$, where $\ket{0}_y$ and $\ket{1}_y$ denote the empty and occupied state, respectively. In the limit of purifying all of the sites, only a
single sample would be required. This limit is known as the ``purification
method'', and while requiring no sampling, it has a rather high cost in the low-temperature limit. 

The optimal choice of which and how many sites to purify turns out to be
delicate and model dependent, as detailed in
Refs.~\onlinecite{Jing2020,Bauernfeind2022}, and one aims to balance a reduction in
the number of samples against an increase in the growth of entanglement. A more detailed discussion on the dependence of the sampling efficiency and bond dimension growth on the number of purified sites, $N_p$, is presented in Appendix ~\ref{app:purification}. In
this work, we choose to purify the bath sites that fluctuate most by purifying
the first $N_p$ bath sites which have the lowest absolute on-site potential
$|\epsilon^i|$, while keeping impurity degrees of freedom unpurified. With this
purification scheme, we anticipate sampling primarily the impurity degrees of
freedom, as fluctuations related to the bath degrees of freedom are mainly
addressed through purification.

\subsection{Improved Estimator}\label{sec:result:improved_estimator} 

By generating a sufficiently large number of $N_S$ samples through Monte Carlo sampling as
described above, the Green's function $G_{\alpha_1\alpha_2}(\tau)$ in
Eq.~\eqref{eq:singl_particle} can be computed as the average of estimators measured on each sample $i$ as 
\begin{align}\label{eq:original_estimator}
&G_{\alpha_1\alpha_2}(\tau) = \frac{1}{N_S}\sum_{i=1}^{N_S}
\begin{cases}
&\estG_{\alpha_1\alpha_2}^{i>}(\tau), \text{for $0\leq \tau\leq\beta/2$ }, \\ \nonumber
&\estG_{\alpha_1\alpha_2}^{i<}(\tau),\text{for $\beta/2 \leq\tau\leq\beta$ }, \nonumber
\end{cases}\\
&\estG_{\alpha_1\alpha_2}^{i>}(\tau) \equiv -\bra{\phi_i} \ophat{d}{\alpha_1}(\tau)\ophat{d}{\alpha_2}^\dagger\ket{\phi_i},\\ \nonumber
&\estG_{\alpha_1\alpha_2}^{i<}(\tau) \equiv -\bra{\phi_i} \ophat{d}{\alpha_2}^\dagger\ophat{d}{\alpha_1}(\tau-\beta)\ket{\phi_i}.
\end{align}
Here, $\estG_{\alpha_1\alpha_2}^{i>}(\tau)$ and
$\estG_{\alpha_1\alpha_2}^{i<}(\tau)$ are referred to as the greater and lesser
estimator, respectively. The splitting of the estimator into $\estG_{\alpha_1\alpha_2}^{i>}(\tau)$ for
$\tau\leq\beta/2$ and $\estG_{\alpha_1\alpha_2}^{i<}(\tau)$ for $\tau>\beta/2$
serves to prevent the overflow of the norm during the imaginary time evolution
and reduce the computational cost by reusing states generated during
calculating $\ket{\phi_i}$~\cite{Bauernfeind2022}.

However, as observed in Ref.~\onlinecite{Bauernfeind2022} and as shown in Fig.~\ref{fig:result:oneband:improved_estimator}(a)(c), 
the above estimators exhibit significant variance as $\tau$ approaches $\beta/2$.
This large variance arises from a mismatched importance sampling between the METTS states $\ket{\phi_i}$
sampled at a temperature $\beta$ and the overlaps of the states \mbox{$\big[\!\bra{\phi_i} \ophat{d}{\alpha_1}(\tau) \big] \big[ \ophat{d}{\alpha_2}^\dagger\ket{\phi_i} \big]$} or \mbox{$\big[\!\bra{\phi_i} \ophat{d}{\alpha_2}^\dagger \big] \big[ \ophat{d}{\alpha_1}(\tau-\beta)\ket{\phi_i} \big]$}
which are not necessarily small when the probabilities $p(i)$ are small.

%Specifically, taking $\tau=\beta/2$ in e.g. $g^>$,  we have 
%\begin{align}
%&\estG_{\alpha_1\alpha_2}^{i>}(\beta/2) = -\underbrace{\bra{i}
%\ophat{d}{\alpha_1}}_{\bra{\psi_i}} \underbrace{ e^{-\beta\hat{H}/2}
%\ophat{d}{\alpha_2}^\dagger \ket{\phi_i}}_{\ket{\varphi_i}} /\sqrt{p(i)}. %\\
%%&\estG_{\alpha_1\alpha_2}(\beta/2)^{i-} = -\bra{\phi_i}
%%\ophat{d}{\alpha_2}^\dagger e^{-\beta/2\hat{H}} \ophat{d}{\alpha_1}  \ket{i}.
%\end{align}  
%For large $\beta$, the state $\ket{\varphi_i} \equiv e^{-\beta\hat{H}/2} \ophat{d}{\alpha_2}^\dagger \ket{\phi_i}$ is close to the ground state (of the sector with $N+1$ particles if the state $\ket{i}$ has $N$ particles). Meanwhile, $\ket{\psi_i}=\ophat{d}{\alpha_1}^\dagger \ket{i}$ is a product state. Consequently, as observed in~\cite{Bauernfeind2022}, the greater estimator $\estG_{\alpha_1\alpha_2}^{i>}(\beta/2) \propto \braket{\psi_i}{\varphi_i}$ is proportional to the overlap between a highly entangled state and a product state, thus it exhibits large fluctuations from sample to sample. One way to mitigate the substantial statistical variance around $\beta/2$ is to purify the impurity site, which increases the overlap $\braket{\psi_i}{\varphi_i}$, but at the cost of considerably increased bond dimensions for accurately describing the many-body state evolved to the same $\beta/2$. 

In order to mitigate this issue, 
we rewrite the decomposition of $G$ differently. Using the notation 
$\tilde{\tau}=\beta-\tau $, we introduce {\it improved estimators} $\bar g$
such that
%To circumvent this problem, note that the overlap $\braket{\psi_i}{\varphi_i}$
%increases significantly \OP{really ? why ? why important? skip this?}
%has a weaker dependence on the initial state
%$\ket{i}$ if $\ket{\psi_i}$ undergoes some amount of imaginary time evolution.
%To take advantage of this fact, we introduce the following improved estimator for the greater and lesser component as
\begin{align}\label{eq:improved_estimator}
&G_{\alpha_1\alpha_2}(\tau)  = \frac{1}{N_S}\sum_{i=1}^{N_S}
\begin{cases}
&\bar\estG_{\alpha_1\alpha_2}^{i>}(\tau), \text{for $0\leq \tau\leq\beta/2$ }, \\ \nonumber
&\bar\estG_{\alpha_1\alpha_2}^{i<}(\tilde{\tau}),\text{for } \beta/2\leq\tau\leq\beta\nonumber
\end{cases}\\
&\bar \estG_{\alpha_1\alpha_2}^{i>}(\tau) \equiv -\bra{\phi_i^{(\beta-\tau)}} \ophat{d}{\alpha_1}{}e^{-\frac{\tau\ophat{H}{}}{2}}
e^{-\frac{\tau\ophat{H}{}}{2}}\ophat{d}{\alpha_2}^\dagger\ket{\phi_i^{(\beta-\tau)}},\\ \nonumber
&\bar \estG_{\alpha_1\alpha_2}^{i<}(\tilde{\tau}) \equiv -\bra{\phi_i^{(\beta-\tilde{\tau})}} \ophat{d}{\alpha_2}^\dagger e^{-\frac{\tilde{\tau}\ophat{H}{}}{2}}
e^{-\frac{\tilde{\tau}\ophat{H}{}}{2}}\ophat{d}{\alpha_1}\ket{\phi_i^{(\beta-\tilde{\tau})}}, \\ \nonumber
&\ket{\phi_i^{(\tau)}} \equiv \frac{e^{-\frac{\tau\ophat{H}{}}{2}}\ket{i}}{\sqrt{\bra{i} e^{-\beta\ophat{H}{}} \ket{i}}}.
\end{align}
We provide evidence in the next section that these improved estimators 
strongly reduce the variance %do not suffer from a large variance
problem near $\tau=\beta/2$. 
%This expression does not suffer from the previous issue, since the value of the improved estimator at each temporal point $\tau$ is determined through the overlap of two states that are both propagated to a total time $\beta/2$. 
%We therefore expect that the variance at long imaginary time will be smaller and more controlled, as will be demonstrated in the next section.
They also has a natural particle hole symmetry.
%This
%approach is analogous to the symmetric form of the METTS method for measuring
%static quantities, offering the advantage of reducing variance at every $\tau$
%point compared to the original estimator. 
Furthermore, the maximal evolved
time in this new scheme is $\beta/2$, as opposed to $\beta$ of the original
estimator—a difference that might be computationally significant for models with highly entangled low-temperature or ground state physics. 
The drawback of the improved estimators is that they require one
to compute $\estG_{\alpha_1\alpha_2}^{i\lessgtr}(\tau)$ at each
$\tau$ point separately. Nevertheless, this issue is strongly mitigated by the use of 
the compact discrete Lehmann representation (DLR) of the Green's function~\cite{Jason2022}. 
The DLR approach allows the Green's function on the whole imaginary time grid to
be efficiently represented to high precision by merely computing its values on a modest
number of special $\tau$ points making up the ``DLR grid'', whose number grows only logarithmically with the inverse temperature.

%%%%%%%%%%%%%%%%%%%%%%%%%%%%%%%%%%%%%%%%%%%%%%%%%%%%%%%%%
%%%%%%%%%%%% BENCHMARKS 
%%%%%%%%%%%%%%%%%%%%%%%%%%%%%%%%%%%%%%%%%%%%%%

\section{Result and Discussion}\label{sec:results}

In this section, we first present a benchmark of the improved estimator
and then apply the METTS algorithm method to the archetypical Hund's metal \sro. 
The appendices present additionnal material, such as: detailed discussions on the
number of bath sites and purified sites are presented in Appendix
~\ref{app:bathsize} and ~\ref{app:purification}; benchmark of the method
against the continuous-time hybridization expansion quantum Monte Carlo algorithm (CTHYB) 
for a three-band Kanamori model
in Appendix \ref{app:threeband}, with an excellent agreement of the
self-energy down to very low temperature $\beta D= 800$ (where $D$ is the
half-bandwidth).

\subsection{Improved Estimator}\label{sec:result:improved_estimator}

We first investigate the effect of the improved estimator on the 
large fluctuations observed in the previous METTS computation around $\tau = \beta/2$ \cite{Bauernfeind2022}.
We consider the DMFT solution of a single-band Hubbard model on the Bethe lattice with a filling of
$n=0.8$ and an interaction strength of $U=4D$, where $D$ denotes the half-bandwidth of the semielliptic density of states. 
\begin{figure}[tb]
\centering
  \includegraphics[width=\linewidth]{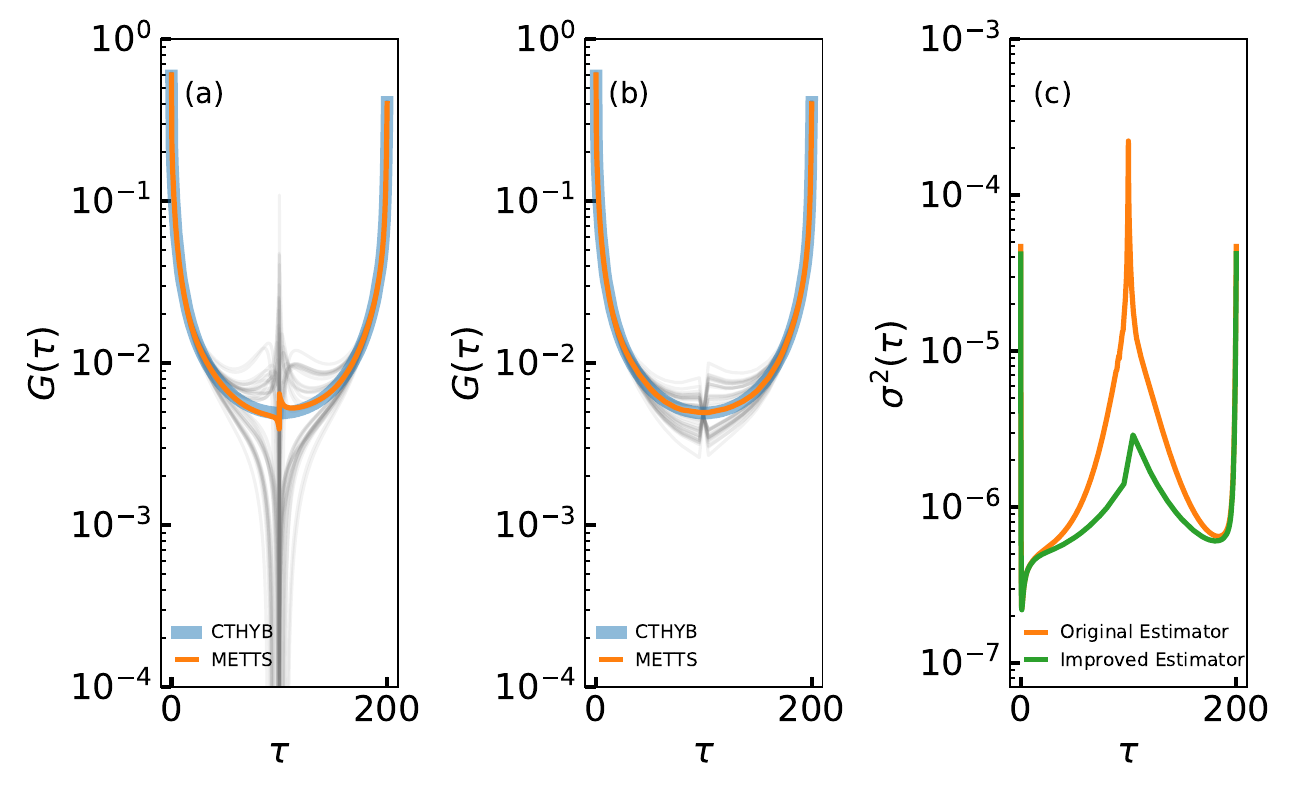}
  \caption{\label{fig:result:oneband:improved_estimator} 
The single-particle Green's function of DMFT solution of the single-orbital Hubbard
model from (a) the original estimator and (b) the improved estimator at $\beta D=400$. The cyan line is the CTHYB  quantum Monte Carlo result, using the implementation Ref.~\onlinecite{Seth2016}, and the grey lines are the estimators measured on each sample.
(c) The METTS variance defined in Eq.~\eqref{eq:defvarianceG} of the original estimator (orange line) and improved estimator (green line). In both cases, $N_S=240, N_p = 4$ and $N_b=8$.}
\end{figure}

In Fig.~\ref{fig:result:oneband:improved_estimator}, we show the Green's function as a function of imaginary time $\tau$ measured from the original estimator on panel {\it a)} and improved estimator on panel {\it b)}.
When employing the original estimator, our benchmarks are in excellent agreement with the CTHYB quantum Monte Carlo results using the implementation in the TRIQS software~\cite{TRIQS2015, Seth2016}
across the entire imaginary time domain, except for \mbox{$\tau \approx \beta/2$}
due to large fluctuations in the METTS sampling, as discussed previously~\cite{Bauernfeind2022}. 
As shown in Fig.~\ref{fig:result:oneband:improved_estimator}(b), the large variance
close to $\beta/2$ is strongly reduced with the improved estimator. 
%Furthermore, by inspecting the sampled Green's functions
%(grey lines), we find the improved estimator is more regularized than the
%original one.
In order to quantify this effect, we consider the METTS variance defined as
\begin{align}\label{eq:defvarianceG}
    &\sigma^2(\tau) \equiv \frac{1}{N_S}\sum_{i=1}^{N_S} \lVert \estG^i(\tau) - G(\tau) \rVert^2.
\end{align}
where $\estG$ is the estimator and $G$ its average over the $N_S$ samples.
%As explained above, the result at convergence ($N_S=\infty$) is independent of the choosing of specific estimator, 
%but the fluctuations at finite $N_S$ are not.
In Fig.~\ref{fig:result:oneband:improved_estimator}(c), we indeed see that the improved estimator significantly reduces the variance around $\beta/2$. 

%-------------------------------------------------------

\subsection{Application to \sro }\label{sec:result:sro}

\begin{figure*}[tb]
\centering
  \includegraphics[width=\textwidth]{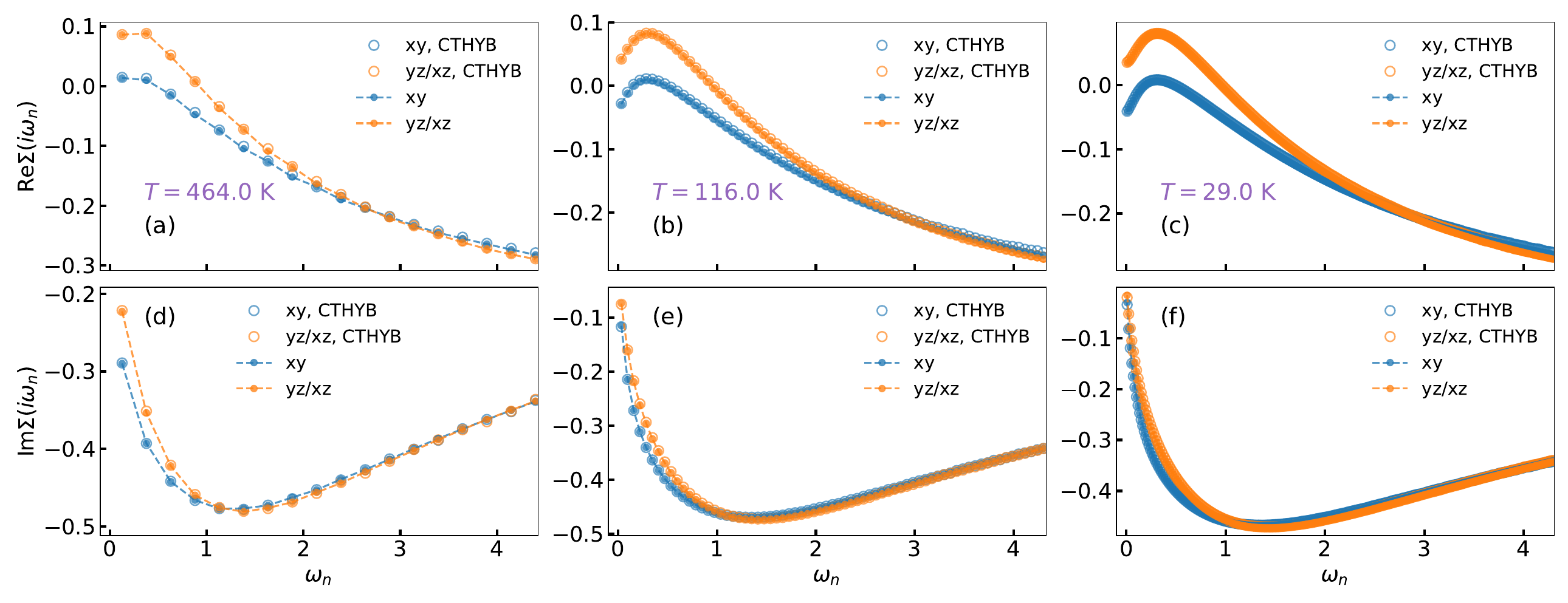}
  \caption{\label{fig:SRO:nosoc} Comparison of the real (a-c) and imaginary (d-f) parts of the DMFT self-energy of \sro without spin-orbit coupling, calculated using the METTS solver (full circles) and CTHYB quantum Monte Carlo (open circles) at $T=464$ K (left panels), $T=116$ K (middle panels), and $T=29$ K (right panels).}
\end{figure*}

Let us now consider \sro~\cite{Mackenzie2003}, within the DMFT framework. 
In addition to the still-debated unconventional superconducting state below $\sim1.5$
K~\cite{maeno1994superconductivity,mackenzie2017even}, the normal state,
including the Hund's metal state~\cite{Georges2013,deng2019signatures} and
Fermi liquid state below $T_{\text{FL}}\approx 25$
K~\cite{Mackenzie2003,kugler2020strongly}, has attracted considerable
attention. Previous DMFT studies have been successfully applied to explain
various experimental observations, such as mass
enhancement~\cite{mravlje2011coherence, tamai2019high}, static magnetic
responses~\cite{strand2019magnetic}, and the enhancement of spin-orbital
coupling strength by a factor of two due to correlation
effects~\cite{kim2018spin,linden2020}. 
Due to the availability of results with multiple methods (CTHYB, NRG), 
this material is an excellent benchmark for multiorbital quantum impurity solvers in a realistic DMFT setup.

The low-energy physics of \sro is determined by the Ru $4d$-$t_{2g}$ orbitals hybridizing with oxygen $2p$ orbitals~\cite{haverkort2008strong,Georges2013}. A minimal one-body Hamiltonian can be constructed by considering the three maximally localized $t_{2g}$ orbitals, which are obtained from the DFT Kohn-Sham orbitals without spin-orbit coupling~\cite{wien2k_a,wien2k_b,wannier90}. The noninteracting density of states of \sro consists of a quasi-two-dimensional $d_{xy}$ band and two degenerate quasi-one-dimensional $d_{yz/xz}$ bands. Notably, the $d_{xy}$ band exhibits a van Hove singularity slightly above the Fermi energy, while its spectral function features a long tail below the Fermi energy and a relatively larger bandwidth than the $d_{xz/yz}$ bands. The latter bands display typical quasi-one-dimensional characteristics with singularities located at their edges~\cite{haverkort2008strong}. 
The Coulomb interaction takes the Kanamori form as
\begin{align}\label{eq:kanamori}
  &\hat{H}_\mathrm{K} = \frac{1}{2}(U-3J) \hat{\mathbf{N}}(\hat{\mathbf{N}}-1)+\frac{5}{2}J\hat{\mathbf{N}}-2J\hat{\mathbf{S}}^2 - \frac{1}{2}J\hat{\mathbf{L}}^2,
\end{align}
with $U=2.3$ eV and $J=0.4$ eV ~\cite{zhang2016fermi,kugler2020strongly,kim2018spin}. Here $\hat{\mathbf{N}}$, $\hat{\mathbf{L}}$, and $\hat{\mathbf{S}}$ are the total particle number, orbital, and spin momentum operators of the impurity, respectively.
The SOC term in the Hamiltonian reads
\begin{equation}\label{eq:soc}
  \ophattext{H}{soc}=\mathrm{i} \frac{\lambda}{2} \sum_{m m' m''}\zeta_{m m' m''} \sum_{\sigma\sigma'} d^\dag_{m\sigma} d_{m'\sigma'} \tau^{m''}_{\sigma\sigma'},
\end{equation}
where $\vb*{\zeta}$ is the completely antisymmetric tensor and $\vb*{\tau}$ the Pauli vector. An isotropic coupling constant of $\lambda= 0.11$ eV is chosen for this term, with $m \in \{d_{xy}, d_{yz}, d_{xz} \}$ and $\sigma \in \{\uparrow, \downarrow \}$.
We use $N_b=8$, $N_p=4$ for calculations without SOC, 
and $N_b=6$, $N_p=2$ with SOC. In every calculation, the measurements are averaged
over $N_S=1080$ samples, with each sample calculated using a maximal bond
dimension of $m=240$ and a truncation error cutoff of $t_w=10^{-10}$. Due to the
local $D_{4h}$ point group of Ru atoms, we omit the spin index when presenting
results without SOC.

We start with results without SOC. In Fig.~\ref{fig:SRO:nosoc}, we compare the
real (upper panels) and imaginary (lower panels) parts of the DMFT self-energy
obtained from our approach with CTHYB quantum Monte Carlo at three temperatures, 
located in the fully incoherent ($T=464$K), crossover
($T=116$K), and coherent ($T=29$K) regions~\cite{mravlje2011coherence}. 
We observe excellent agreement between our results and CTHYB at these three
typical temperatures, demonstrating the accuracy of our method.

\begin{figure}[tb]
\centering
  \includegraphics[width=\linewidth]{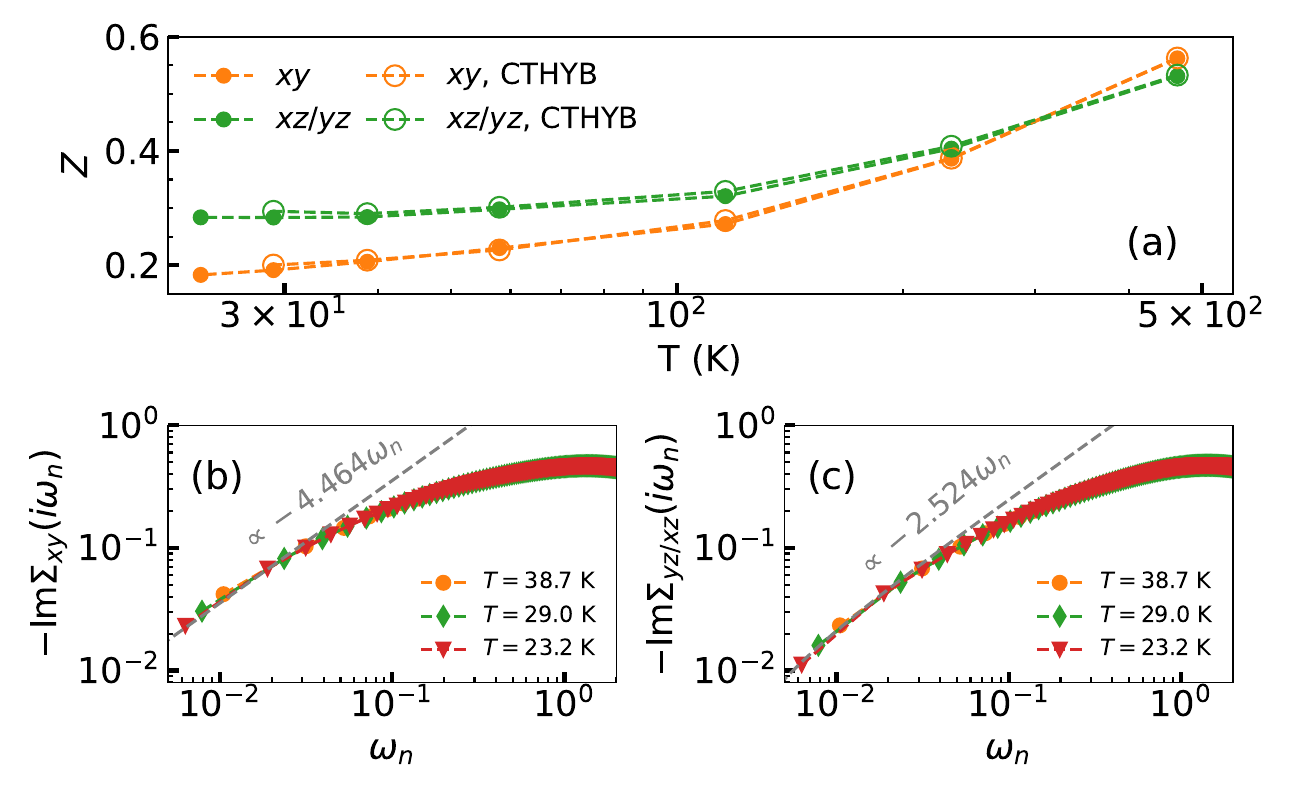}
  \caption{\label{fig:result:SRO:Z} (a) Comparison of quasiparticle weights $Z$ from our approach (filled circles) and CTHYB quantum Monte Carlo (open circles) at various temperatures. (b) Imaginary part of the DMFT self-energy for the $d_{xy}$ orbital and (c) the two degenerate $d_{yz}$ and $d_{xz}$ orbitals at temperatures of $T=38.7$ K (orange circles), $T=29$ K (green diamonds), and $T=23.2$ K (red triangles). The dashed grey lines serve as guides to the eye and are linear with respect to $\omega_n$.}
\end{figure}
Our approach's capability to provide accurate results across large temperature range is further demonstrated by comparing the extracted quasiparticle weight $Z=(1-\partial_{\omega_{n}} \mathrm{Im} \Sigma(i\omega_n)|_{\omega_n\rightarrow 0})^{-1}$ from our method with CTHYB in Fig.~\ref{fig:result:SRO:Z}(a)
\footnote{
The quasiparticle weight $Z$ is determined by fitting a polynomial of 4th order to the lowest 6 points of the Matsubara self-energies.
}.
We observe excellent agreement for temperatures as high as $T=464$ K down to
low temperatures as $T=29$ K, which is close to the Fermi-liquid temperature
$T_{\text{FL}}\approx 25$ K~\cite{Mackenzie2003,kugler2020strongly}. Notably,
accurately calculating the first few Matsubara frequencies of the self-energy
can be challenging for CTHYB at very low temperatures. Our method, however,
benefiting from fork tensor product state representation of the impurity wave function, is less
constrained by low-temperature calculations %(scales as $\beta m^4$) 
and can
access even lower one as $T=23.2$ K.

Fig.~\ref{fig:result:SRO:Z}(b) and (c) depict the imaginary part of the
self-energy of the $d_{xy}$ and $d_{yz/xz}$ orbitals. For $T=23.2$ K, we
observe a linear behavior of $\mathrm{Im}\Sigma(i\omega_n)$ as
$\omega_n\rightarrow 0$ for all three orbitals, indicating Fermi liquid
behavior. The quasiparticle weight extracted at $T=23.2$ K is $0.18$ %$0.183
for the $d_{xy}$ orbital and $0.28$ %$0.284$ 
for the $d_{yz/xz}$ orbitals, in good agreement
with previous calculations~\cite{kugler2020strongly,kim2018spin} and
experimental results~\cite{Mackenzie2003,tamai2019high}. Despite the larger
bandwidth of the $d_{xy}$ orbital, its quasiparticle excitations are more
renormalized than the $d_{yz/xz}$ orbitals. This unexpected behavior can be
attributed to the relatively smaller spectral weight of the hybridization
function for the $d_{xy}$ orbital near the Fermi energy due to the van Hove
singularity~\cite{mravlje2011coherence}.

\begin{figure}[tb]
\centering
  \includegraphics[width=\linewidth]{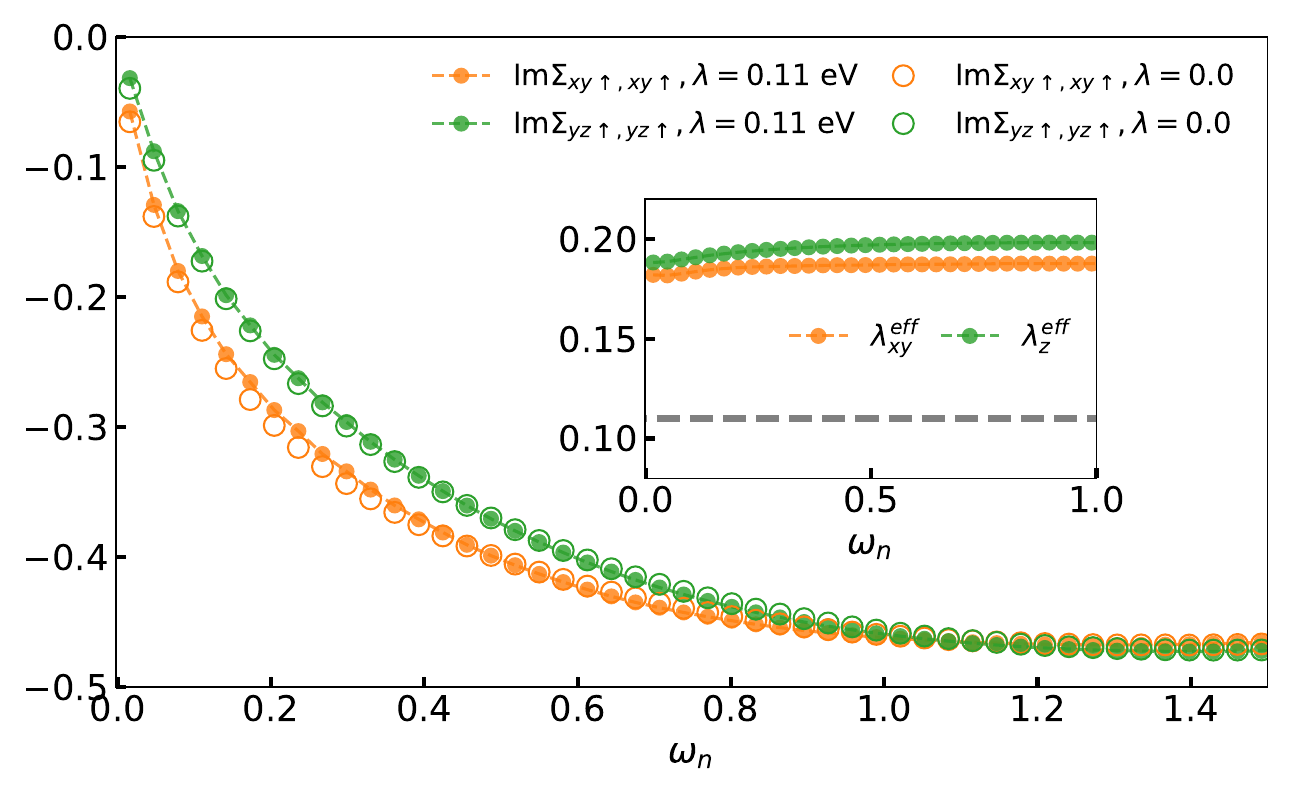}
\caption{\label{fig:SRO:SOC} Comparison of the imaginary part of selected self-energy components for \sro with (filled circles) and without (open circles) spin-orbit coupling at $T=58$ K. The inset illustrates the effective spin-orbit coupling strength $\lambda_{xy}^{\text{eff}}$ (orange) and $\lambda_{z}^{\text{eff}}$ (green). The grey dashed line represents the bare coupling strength of $\lambda=0.11$ eV.}
\end{figure}
Let us now turn to computations with SOC. This term is known to play a
significant role in this material, although it is only a few percent of the
bandwidth, its inclusion is crucial for accurately describing
the experimental Fermi
surface~\cite{tamai2019high,haverkort2008strong,Cao2021}. Incorporating SOC
is particularly challenging for QMC solvers, with calculations performed only
for temperatures above 200 K~\cite{kim2018spin,zhang2016fermi,karp2020sr}.
While zero-temperature results are available from MPS-based solvers, using an
artificial inverse temperature of $\beta^{\text{eff}}=200$
eV$^{-1}$~\cite{linden2020}, the impact of SOC at intermediate temperatures on
single-particle quantities remains to be explored with a direct computation.
Here, we present results with SOC at $T=58$ K.
In Fig.~\ref{fig:SRO:SOC}, we compare $\mathrm{Im}\Sigma(i\omega_n)$  with (filled circles) and without (open circles) SOC at $T=58$ K. As anticipated, the diagonal elements are only slightly modified, and the quasiparticle weights slightly change from $0.226$ to $0.228$ for the $d_{xy}$ orbital and from $0.301$ to $0.306$ for the $d_{yz/xz}$ orbitals. As depicted in the inset, the effective
in and out-of-plane spin-orbital coupling strengths respectively defined as
\begin{subequations}
\begin{align}
   \lambda_{xy}^{\text{eff}}(i\omega_n)&\equiv\lambda -2\mathrm{Im}\Sigma_{xy\uparrow,xz\downarrow}(i\omega_n),
   \\
   \lambda_{z}^{\text{eff}}(i\omega_n)&\equiv\lambda +2\mathrm{Im}\Sigma_{yz\uparrow,xz\uparrow}(i\omega_n)
\end{align}
\end{subequations}
are essentially frequency-independent over a wide frequency range and can be
considered as constant single-particle terms added to the bare ones. This
enhancement of SOC strength by correlation effects, resulting in
$\lambda_{xy}^{\text{eff}}\approx 0.19$ %0.188
eV and
$\lambda_{z}^{\text{eff}}\approx 0.2$ %0.198
eV, lifts the degeneracies at the
crossing along the Brillouin zone and gives rise to well-separated sheets,
consistent with experimental observations~\cite{tamai2019high}. The originally
isotropic atomic SOC now exhibits some anisotropy between the in- and
out-of-plane values due to the lower-symmetry local point group $D_{4h}$ of the
Ru ion. These findings are in agreement with earlier finite temperature CTHYB
calculations ~\cite{kim2018spin,sro214_qp_hightem} and zero-temperature MPS calculations~\cite{kugler2020strongly,linden2020}.

%%%%%%%%%%%%%%%%%%%%%%%%%%%%%%%%%%%%%%%%%%%%%%%%%%%%

\section{Conclusion}
\label{sec:conclusion}

In summary, we have introduced an improved estimator
for Green's function in METTS computations. Using this estimator significantly reduces 
the variance in certain time ranges where the previous estimator of Ref.~\cite{Bauernfeind2022} would require 
many more samples.
Combining the METTS and purification approach with the fork tensor product state network, we obtained
a finite-temperature quantum impurity solver which works
for three-band systems at low temperature, even in the presence of spin-orbit coupling,
a regime inaccessible to the quantum Monte Carlo methods.
We illustrated our method with a computation of the self-energy for \sro\ 
including the spin-orbit coupling.

Because both the improved $G(\tau)$ estimator and the METTS algorithm are quite general purpose, we expect the above improvements to be useful for studying strongly-correlated lattice models where METTS has already been used to compute thermodynamic properties \cite{WietekStripes2021,WietekMott2021,Feng2023}.
Having controlled access to the imaginary-time Green's function should help to understand scenarios such as the
disordered intermediate-$U$ phase of the triangular lattice Hubbard model \cite{WietekMott2021}.

%%%%%%%%%%%%%%%%%%%%%%%%%%%%%%%%%%%%%%%%%%%%%%%%%%%%

\begin{acknowledgments}
The Flatiron Institute is a division of the Simons Foundation.
\end{acknowledgments}
%%%%%%%%%%%%%%%%%%%%%%%%%%%%%%%%%%%%%%%%%%%%%%%%%%%%%%%%%
%%%%%%%%%%%  APPENDICES
%%%%%%%%%%%%%%%%%%%%%%%%%%%%%%%%%%%%%%%%%%%%%%%%%%%%%%%%%

\appendix
\counterwithin{figure}{section}

\section{Fit of the hybridization function}\label{app:DeltaFit}

In this section, we give the details of our fitting procedure of the hybridization function
to the form Eq.\eqref{eq:finitebathDelta}.
For a given  hybridization function $\Delta$, the parameters $\epsilon$ and $V$ are  determined by minimizing a cost 
function~\cite{Georges1996}
\begin{equation}
   \chi(\{\epsilon, V\}) = \frac{1}{N_{\text{fit}}}\sum_{n=0}^{N_\text{fit}} \frac{1}{\omega_n^\gamma} \lVert \Delta(i\omega_n) - \bar\Delta(i\omega_n)\rVert,
\end{equation}
where $\omega_n$ are Matsubara frequencies, $N_\text{fit}= 1500$ in this paper and $\lVert\cdot\rVert$ is the Frobenius norm. 
The parameter $\gamma\in[0,1]$ adjusts the weighting of low frequencies in the
cost function $\chi$. For $\gamma=0$, all frequencies are treated equally,
while low frequencies are assigned greater weight for $\gamma>0$. We found that
choosing $\gamma=1$ stabilizes our DMFT convergence, particularly in cases
where a small number of bath sites $N_b$ is used.

%%%%%%%%%%%%%%%%%%%%%%%%%%%%%%%%%%

\section{Implementation and Computational details}\label{app:implementation_details}
In this appendix, we elaborate on the implementation and computational details of the METTS solver. To calculate each $\estG_{\alpha_1\alpha_2}(\tau)$, one first evolves a product state $\ket{i}$ in imaginary time to $\ket{\phi_i} = e^{-\beta\hat{H}/2} \ket{i}/\sqrt{\bra{i} e^{-\beta\hat{H}} \ket{i} }$. Due to the non-local nature of the impurity Hamiltonian and the absence of direct interaction between the bath degrees of freedom, the two-site TDVP cannot adjust the bond dimension during an imaginary time evolution. Therefore, one can either employ an alternative time evolution method, such as the global Krylov time evolution method, during the initial time steps, and then switch to either two-site or single-site TDVP up to $\beta/2$. In our implementation, we extend the global basis expansion method proposed in~\cite{Mingru2020} for MPS to fork tensor product states. In each of the time evolution steps, $\ket{\psi(\tau+\Delta\tau)} = e^{-\Delta\tau\hat{H}}\ket{\psi(\tau)}$, a series of $N_k$ reference states are generated as
\begin{align}
\left\{ \left(1-\Delta\tau\hat{H}\right)\ket{\psi(\tau)},\cdots,\left(1-\Delta\tau\hat{H}\right)^k\ket{\psi(\tau)}\right\}, \nonumber
\end{align}
and used to expand the basis of $\ket{\psi(\tau)}$. In general, the sequential application of $\left(1-\Delta\tau\hat{H}\right)$ onto an fork tensor product state increases the bond dimension rapidly and is computationally expensive. However, since high accuracy is not required for each reference state for basis expansion, we truncate these reference states with a relatively small bond dimension of around $m'=50$. Moreover, we find that in all our calculations, a relatively small $N_k=1,2$ is typically sufficient.

For calculating each METTS state, we employ an exponential growth imaginary time grid as $\{\tau_0\Delta\tau^0, \tau_0\Delta\tau^1, \tau_0\Delta\tau^2,\cdots,\beta/2\}$, which ensures a small time step at the beginning of the imaginary time evolution, while increasing the time steps as approaching $\beta/2$. Such an exponential time grid has been demonstrated to speed up calculations exponentially and provide better accuracy~\cite{Binbin2018}. In our implementation, we have $\tau_0=0.05$, and $\Delta\tau$ is determined by a given number of imaginary time points $N_\tau$, which is set according to $\beta$. For instance, for the \sro\ calculations, we have $N_\tau=4\beta/5$.

%To measure the estimator $\estG_{\alpha_1\alpha_2}(\tau)$ on each METTS, one needs to further evolve $e^{-\tau\hat{H}}\ophat{d}{\alpha_1}\ket{\phi_i}$ and $e^{-\tau'\hat{H}}\ophat{d}{\alpha_2}^\dagger\ket{\phi_i}$ in imaginary time for $\tau,\tau'\in[0,\beta/2]$. This measurement is computationally demanding and takes up around 60\% of the computation time. Since the Green's function changes most rapidly for $\tau$ around $0$ and $\beta$, determined by the high energy scales, and more mildly around $\beta/2$, determined by the low energy scales, we employ a dense grid for $\tau$ around $0$ and $\beta$, and a sparser grid for $\tau$ around $\beta/2$ (using a time step 10 times larger than those around $0$ and $\beta$). Subsequently, $G(i\omega_n)$ is obtained through a first discrete Lehmann representation (DLR) fit of $G(\tau)$ and transformed into Matsubara frequency with the DLR coefficients~\cite{Jason2022,Jason2002libdlr}.

\section{Bath size dependence }\label{app:bathsize}
\begin{figure}[tb]
\centering
  \includegraphics[width=\linewidth]{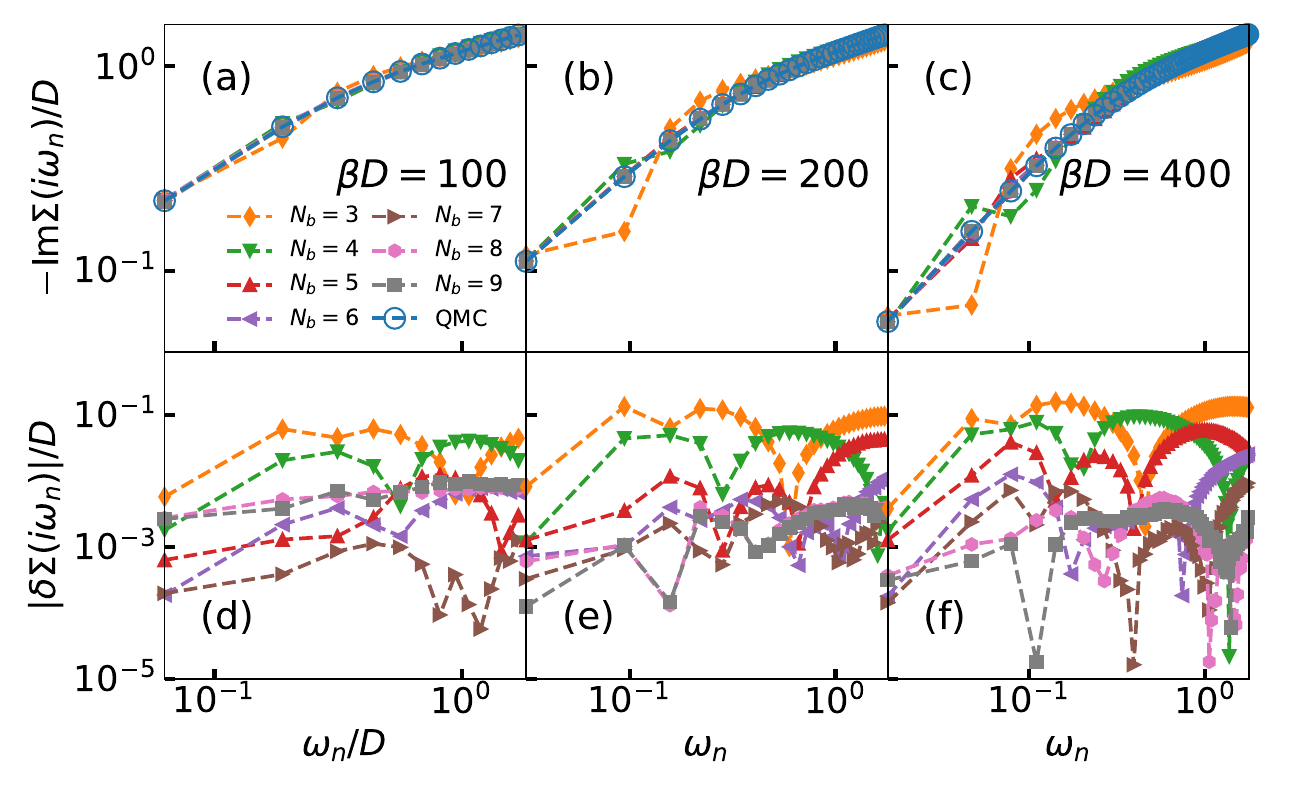}
  \caption{\label{fig:result:oneband:Nb} (a)-(c) Imaginary part of the self-energy, and (d)-(f) corresponding differences compared to QMC results, for DMFT solution of the single-band model at a filling of $n=0.8$ and interaction strength $U=4D$. Results are obtained with varying numbers of bath sites ($N_b=3,4,5,6,7,8$ and $9$) per spin-orbital. Panels are organized by temperature: $\beta D=100.0$ (left), $\beta D=200.0$ (middle), and $\beta D=400.0$ (right).
  }
\end{figure}

Although the convergence of DMFT solutions from an impurity solver employing finite bath sites toward the continuous limit has been extensively discussed \cite{Georges1996,liebsch2011,senechal2010,Zaera2020}, for completeness, we present our solver's results at various temperatures and bath sites in Fig.~\ref{fig:result:oneband:Nb} for the single-band model same as Sec.~\ref{sec:result:improved_estimator} in main text. As $N_b$ ranges from $3$ to $7$, we observe a systematic convergence of $\Sigma(i\omega_n)$ to the QMC continuous limit. Comparing panels (a) and (c), we find that lower temperatures ($\beta D=400$) require larger $N_b\geq 6$ to achieve satisfactory agreement with QMC results, unlike cases at higher temperatures ($\beta D=100$) showing excellent agreement with QMC already with $N_b=4$. For $\beta D=200$ and $400$, increasing $N_b$ further from $8$ to $9$ no longer yields substantial improvements, and for $\beta D=100$, we encounter the "over-fitting" problem for $N_b=8,9$ as indicated by the increase of error $\delta\Sigma(i\omega_n)$ in Fig.~\ref{fig:result:oneband:Nb}(d).

\section{Purification versus sampling. Role of $N_p$}\label{app:purification}

\begin{figure}[tb]
\centering
  \includegraphics[width=\linewidth]{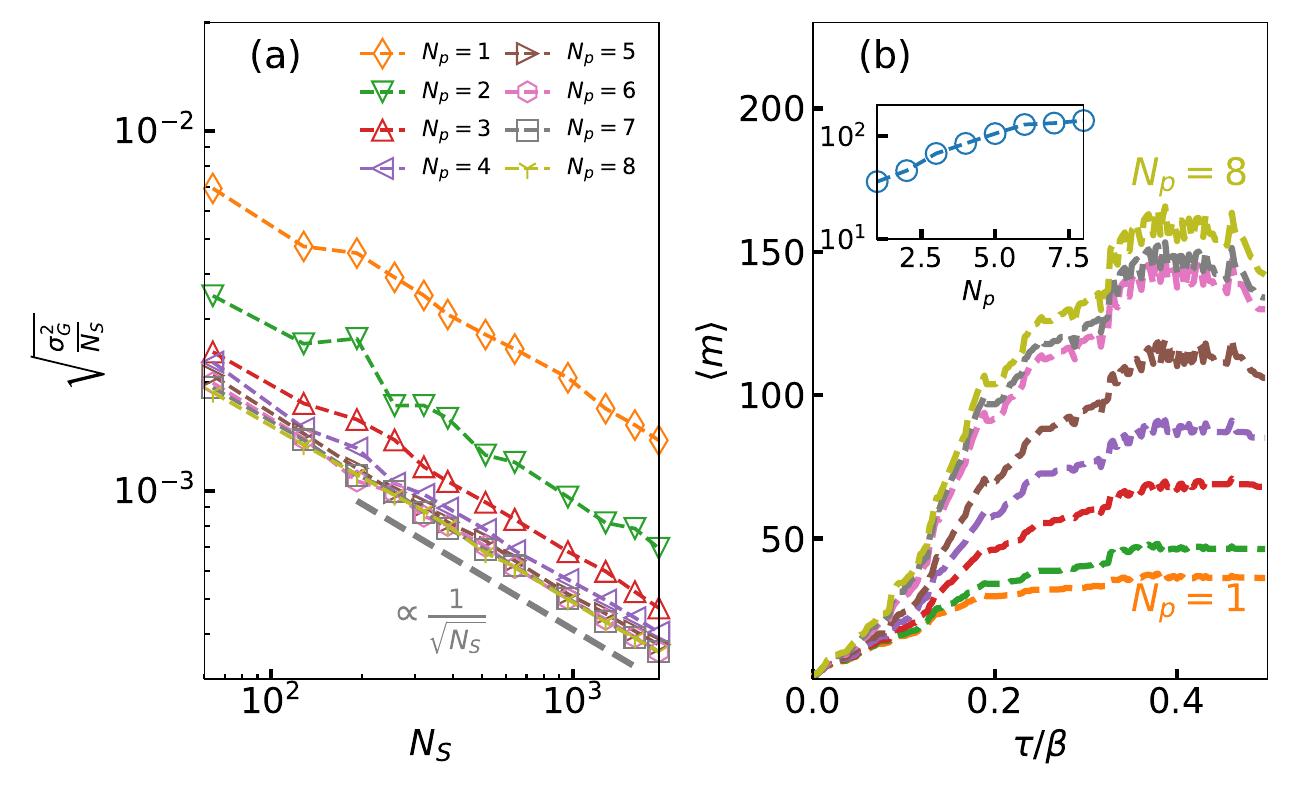}
  \caption{\label{fig:result:oneband:bondAndSample}
(a) Standard error of the mean $\sqrt{\sigma^2_G/ N_S}$ as a function of Monte Carlo sampling size
$N_S$ for various numbers of purified bath sites $N_p$ per spin-orbital in the
DMFT solution of the single-band Hubbard model. The dashed grey line represents
a guide to the eye and is proportional to $1/\sqrt{N_S}$. (b) Average bond
dimension $\langle m\rangle $ with $N_S=120$ Monte Carlo sample, as a function of imaginary time $\tau$ with
various $N_p$ values, and the insert shows $\langle m\rangle $ at
$\tau=\beta/2$ as a function of $N_p$. The calculations are performed with
$N_b=8$ at $\beta D=200$ and a truncation weight of $t_w=10^{-11}$ for the
time evolution.}
\end{figure}

Here we discuss the role of $N_p$, the number of purified sites, using the single-band model same as Sec.~\ref{sec:result:improved_estimator} in the main text.
In Fig.~\ref{fig:result:oneband:bondAndSample}(a), we present the imaginary time
integrated variance 
\begin{equation}
\sigma^2_G = \int_0^\beta \sigma^2(\tau)d\tau
\end{equation}
as a function of the number of samples $N_S$ for increasing $N_p$. 
We see that increasing $N_p$ from $1$ to $4$ strongly reduces the METTS fluctuations.
For $N_p\geq 4$, this reduction is much more modest, and the standard error of the mean, $\sqrt{\sigma_G^2/N_S}$,
starts to follows a clear standard $1/\sqrt{N_S}$ decay rate, characteristic of Monte Carlo sampling.
In  Fig.~\ref{fig:result:oneband:bondAndSample}(b), we show the 
averaged maximal bond dimension $\langle
m\rangle$ versus $\tau$, for various $N_p$. As expected, the bond dimension grows with $N_p$, 
and with imaginary time $\tau$. The growth with $\tau$ tends to saturate, 
due to the fact that $\ket{\phi_i}$ converges to the
ground state in the $\beta\rightarrow\infty$ limit.
We also observe that as $N_p\rightarrow N_b$, the average bond dimension increases more slowly:
the purification of the bath sites far away from Fermi level has little effect, probably due to their fillings being 
very close to $0$ and $1$.
We conclude that, for these parameters, there is actually an optimum in $N_p$ around $N_p \approx 4$ above which 
the computational cost of forming each METTS (measured by the average bond dimension) continues to grow with $N_p$, 
while the METTS variance does not decrease significantly anymore.

%\OPadd{Such behaviors can be understood as follows. When choosing a small number of
%bath sites to purify, the probability distribution $p(i)$ comprises a few
%dominant low energy states with large weights corresponding to initial product
%states $\ket{i}$ with bath sites having positive on-site energy being empty and
%negative energy being occupied, and a large number of high energy states with
%exponentially smaller weights corresponding to initial product states having
%bath sites with positive on-site energy being occupied or negative energy being
%empty. These high energy states, lying in the long tail of the distribution of
%$p(i)$, make sampling very inefficient. As $N_p$ increases, contributions from
%these high-energy states are more included by purification, resulting in an
%efficient sampling of the dominant low energy states only. Hence, increasing
%$N_p$ from $1$ to $4$ improves the sampling efficiency. When $N_p$ is further
%increased, the newly included states through purification have much larger
%increased energies, resulting in exponentially small weights and contribuations
%to the measurement. Consequently, sampling efficiency no longer improves.}

%%%%%%%%%%%%%%%%%%%%%%%%%%%%%%%%%%%%%%%%%%%%%%%%%%%%%%%%%%

\section{Benchmark: three-band model }\label{app:threeband}

\begin{figure*}[htb]
\centering
  \includegraphics[width=\textwidth]{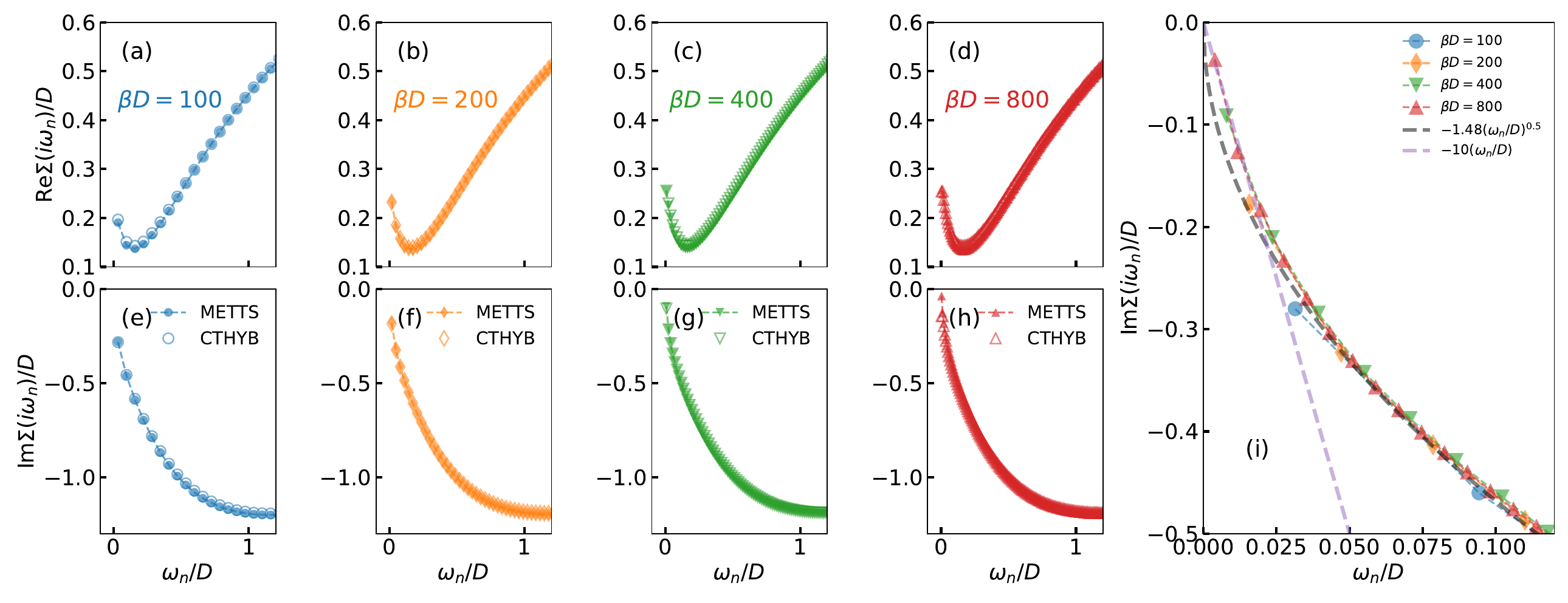}
  \caption{\label{fig:result:threeband:hundmetal_combined} Comparison of (a-d)
  real and (e-h) imaginary part of DMFT self-energy for the degenerate
three-orbital model using the METTS solver (filled symbols) with CTHYB results
(open symbols) at various temperatures. (i) Imaginary part of the self-energy
as a function of Matsubara frequency at various temperatures. Dashed grey and
purple lines indicate the scaling behavior of $\omega_n^{0.5}$ and $\omega_n$,
respectively. For all METTS calculations, we use $N_b=8$ bath sites per
spin-orbital, out of which $N_p=5$ are purified. The number of Monte Carlo
samples is $N_S=500$, and the bond dimension is $m=200$.}
\end{figure*}

In this appendix, we benchmark our approach on a three-band model.
We consider a Kanamori Hamiltonian of form Eq.~\eqref{eq:kanamori} in the main text.
Each impurity orbital is
separately coupled to a bath with an identical semielliptic density of states
of half-bandwidth of $D$. We use a filling of $\langle \hat{N}\rangle=2$, and interaction
strengths of $U=4D$ and $J=U/6$.

In Fig.~\ref{fig:result:threeband:hundmetal_combined}(a)-(h), we see that
the DMFT self-energy $\Sigma(i\omega_n)$ obtained with our method is in very good agreement 
with CTHYB at various temperatures. The precise low-frequency dependency of the self-energy constitutes a 
more rigorous benchmark of the method.
In this "Hund's metal" model, the strong suppression of Kondo
screening scales by Hund's coupling gives rise to a phase characterized by an
almost frozen local spin moment and a power-law behavior in the low-frequency
self-energy, in contrast to ordinary Fermi-liquid behavior~\cite{Werner2008,Werner2009,Georges2013}.
This behaviour is illustrated in
Fig.~\ref{fig:result:threeband:hundmetal_combined}(i): the
self-energy exhibits a $\Sigma(i\omega_n)/D \sim (\omega_n/D)^{0.5}$ power-law behavior at low frequencies. 
For the lower temperature case of $\beta D=800$, a crossover to Fermi-liquid behavior is observed,
indicated by the linear dependence of the self-energy on $\omega_n/D$ at low frequencies.

\begin{figure}[h!]
\centering
  \includegraphics[width=\linewidth]{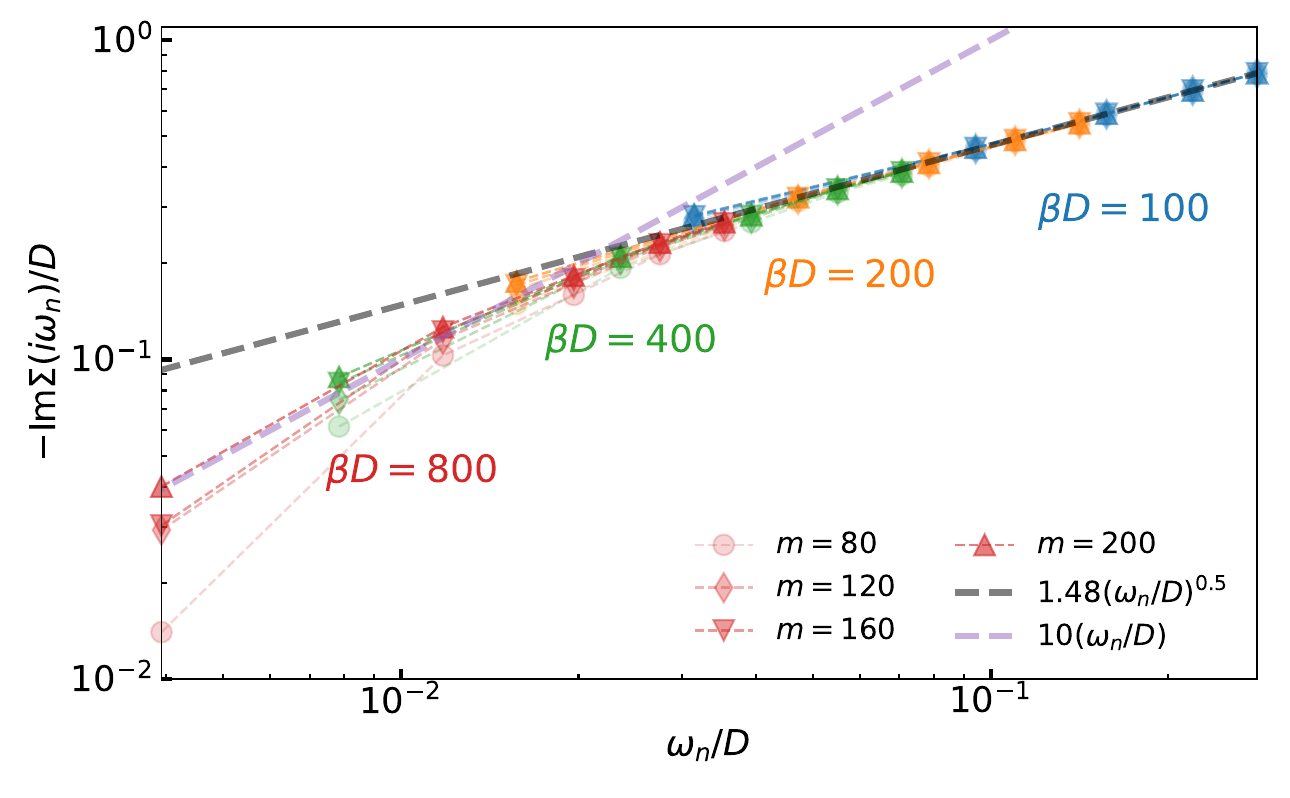}
  \caption{\label{fig:result:threeband:hundmetal_bond} Bond dimension convergence for the degenerate three-orbital model at various temperatures using the same parameters set as Fig.~\ref{fig:result:threeband:hundmetal_combined}. }
\end{figure}

In Fig.~\ref{fig:result:threeband:hundmetal_bond}, we show the convergence of
self-energy with respect to bond dimension at various temperatures for the
three-band model. We observe a systematical convergence of the self-energy with
increasing $m$. The low frequencies are more sensitive to the bond dimension.
For $\beta D=100$, $m=80$ is enough to achieve a satisfactory accuracy, while
for $\beta D=800$, a larger bond dimension of $m=200$ is required.

\bibliographystyle{apsrev4-2}
\bibliography{Solver_ref.bib}

\end{document}